\RequirePackage{fix-cm}
\documentclass[smallextended]{svjour3} 
\usepackage[doublespacing]{setspace}      
\smartqed  
\usepackage{graphicx}
\begin{document}

\title{Structural and magnetic response of CrI$_3$ monolayer to electric field
}


\author{S. Ghosh         \and
        N. Stoji\' c  \and  N. Binggeli}


\institute{Abdus Salam International Centre for Theoretical Physics, Strada Costiera 11, Trieste I-34151, Italy\\
              Tel.: +39-040-2240454\\
              \email{sghosh4@ictp.it}           
}

\date{Received: date / Accepted: date}

\maketitle

\begin{abstract}
A recent theoretical study reported large effects of perpendicular
electric fields on the atomic structure of a monolayer CrI$_3$, which could be
related to the microscopic origin of the technologically promising and
experimentally observed electrical switching of magnetization 
in bilayer CrI$_3$. However, those theoretical results are not in line with a
previous theoretical finding of only slight changes under a strong
electric field in CrI$_3$.  Given the important consequences that the
presence of large structural distortions in an electric field might have,
we investigated the effects of external electric fields on the CrI$_3$ 
monolayer using density functional theory for a wide range of field
strengths.  Conclusively, we find that the structural response of CrI$_3$ to
the applied perpendicular electric field is extremely small due to a very efficient electronic screening of the electric field within the monolayer. Therefore it
cannot be the origin of the observed electrical switching of magnetization 
in the bilayer CrI$_3$. Furthermore, we find that the very small linear
dependence of the structural changes on the electric field persists up to
a field value of 0.45~V/\AA, while the Cr magnetic moment remains constant
for the same strengths of electric field.

\keywords{Chromium trihalide \and electric field \and first-principles calculations \and structural response \and Dzyaloshinskii-Moriya interaction}
\end{abstract}

\section{Introduction}
Electrical control of magnetism is a key challenge in condensed matter physics from not only fundamental aspects but also for technological applications. The materials whose magnetic properties can be controlled by electric field could be useful for low-power and high-speed magnetic switching devices.
Recently discovered two-dimensional magnetic van der Waals semiconductors, like CrI$_3$\cite{HuangNat,C5TC02840J,McGuire}, CrGeTe$_3$\cite{CrGeTe3} and Cr$_2$Ge$_2$Te$_6$\cite{fmsemicond1}, can be incorporated and gated in van der Waals nanostructured devices and are attractive candidates for electrical control of magnetism at the nanoscale. 
In particular, CrI$_3$, which came into focus most recently due to exciting experiments on its bilayer form, \cite{NatMaterJiang,JianNatnanotech,HuangNatNano} is a ferromagnet in bulk with Curie temperature of 61~K with spins pointing out-of-plane\cite{cryst7050121,McGuire,Lado}. Monolayer CrI$_3$ possesses a ferromagnetic (FM) ground state with Curie temperature of 45~K\cite{HuangNat}, while the bilayer is antiferromagnetic (AFM) with opposite magnetic moments from the two FM monolayers below a critical temperature of about 45 K\cite{HuangNatNano}.
Very recently, a large linear magnetoelectric (ME) effect has been evidenced experimentally in the AFM bilayer CrI$_3$, leading to electrically controlled magnetism characterized by a linear dependence of the magnetization in the bilayer ground state\cite{NatMaterJiang}. This was observed with applied perpendicular electric fields up to 0.1V/\AA\cite{NatMaterJiang}. Furthermore, in the presence of a magnetic field near the FM-AFM spin-flip transition, a reversible  electrical switching of the interlayer  magnetic order between the FM and AFM states could be achieved by exploiting the large ME response near the critical field\cite{NatMaterJiang}. The microscopic mechanism responsible for the large ME coupling in the CrI$_3$ bilayer is still unknown.

In recent theoretical studies based on first-principles density functional theory (DFT) calculations, Liu {\it et al.} \cite{chinese_PRB,chinese_skyrmion} predicted very large structural distortions in the CrI$_3$ monolayer in response to vertical electric fields. In particular, in the linear-response regime, splittings due to inversion-symmetry breaking by the electric field in the structural-parameter values of the two surfaces of the CrI$_3$ layer as large as 7 \% were predicted for the I-Cr interlayer distance ($d$), with a field as small as 0.02 V/\AA\cite{chinese_skyrmion}, and as large as 3 \% for the Cr-I-Cr angle ($\theta$), with a field of 0.1 V/\AA\cite{chinese_PRB}. Such large structural distortions were found to give rise to a considerable Dzyaloshinskii-Moriya interaction, leading to an electrical reversal of magnetization\cite{chinese_PRB} and to electric field induced magnetic skyrmions\cite{chinese_skyrmion} in simulations with electric fields smaller than  0.1 V/\AA. Such results would suggest therefore similar large structural changes induced by the electric field in the CrI$_3$ bilayer, which might be related to the large ME response observed in that system.  However, the large structural changes predicted in 
are not in line with a previous first-principles study on the CrX$_3$ (X = Cl, Br, I) monolayer systems in the presence of a much larger electric field of 1 V/\AA, which finds only slight changes induced by the electric field in the band structure of these materials\cite{C5CP04835D}. The actual impact of perpendicular electric fields on the atomic structure of the CrI$_3$ monolayer is thus unclear and is important to know in view of its potential consequences on the Dzyaloshinskii-Moriya interaction and resulting ME effects in the CrI$_3$ monolayer and bilayer systems.

In this study, we systematically examine from first-principles the influence of the external perpendicular electric field on the atomic structure of the CrI$_3$ monolayer. The applied field is varied in a large range to carry out a comprehensive study of the trends with field strength also surpassing the linear-response regime. In contrast to previous theoretical findings, we show that the structural changes induced by the electric field are extremely small and therefore cannot cause any significant Dzyaloshinskii-Moriya effect. The very small amplitude of the structural changes results from a very efficient electronic screening of the electric field within the CrI$_3$ monolayer.

\section{\bf{Computational Details}}
The calculations are performed using spin-polarized {\it{ab initio}} density functional theory (DFT) as implemented in the Quantum ESPRESSO package\cite{QE}. We have used projector augmented wave pseudopotentials \cite{PAW} to describe the electron-ion interactions. The exchange-correlation interactions are treated within the Perdew-Burke-Ernzerhof (PBE) form of the generalized gradient approximation\cite{PBE}. The kinetic energy and charge density cutoffs for the plane wave basis set are chosen to be 48 Ry and 457 Ry, respectively. The periodic images of the monolayer are separated by introducing a vacuum of thickness 40 \AA\ along the $z$-direction. Brillouin Zone is sampled with 8$\times$8$\times$1 $k$-point mesh. The atomic relaxation is done until the force on each atom becomes smaller than 10$^{-4}$ Ry/Bohr. The convergence criterion for electronic self-consistency is set as 10$^{-12}$ Ry in order to accurately capture the correct small structural distortions induced by the applied out-of-plane electric field. The electric field is modeled using a sawlike potential along the direction perpendicular to the plane of the monolayer (the $\hat{z}$ direction). Dipole correction is applied to avoid spurious interactions between the periodic images of CrI$_3$ along the $z$-direction\cite{dipole}.

\section{\bf{Results and Discussion}}
\subsection{\bf{CrI$_3$ structure at zero electric field}}
\begin{figure}[t]
\centering
\vspace{0cm}
{\includegraphics[width=0.5\textwidth]{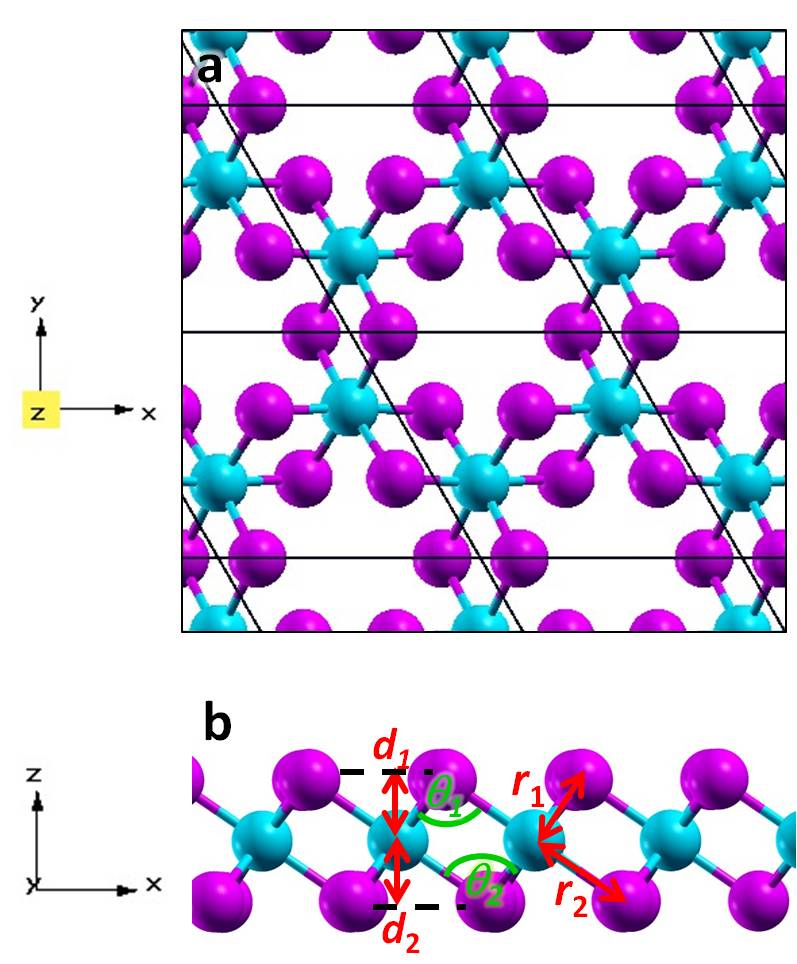}}
\caption{Atomistic structure of CrI$_3$ monolayer. The cyan and magenta show Cr and I atoms respectively. (a) Top and (b) side views of CrI$_3$ monolayer. $d_1$ and $d_2$ are difference between the $z$ coordinates of the Cr and I atoms for the upper and lower halves of the monolayer respectively. $\theta_1$ and $\theta_2$ are the Cr-I-Cr angles for the lower and upper halves of the monolayer. $r_1$ and $r_2$ are the distances between Cr and I atoms in the lower half and upper half of the monolayer respectively.}
\label{CrI3_ml}
\end{figure}

Fig.\ref{CrI3_ml} shows the equilibrium atomic structure of the CrI$_3$ monolayer. 
The unit cell consists of 2 Cr atoms and 6 I atoms as shown by the black lines in Fig.\ref{CrI3_ml} (a). In this structure, each I atom is ionically bonded to two Cr atoms, and each Cr atom is bound to six I atoms (forming a Cr-centered I octahedron). The Cr$^{3+}$ ions  form an hexagonal network in octahedral coordination, edge-sharing with six I$^-$ ions. The CrI$_3$ monolayer consists of three consecutive layers of I, Cr and I atoms as shown in Fig.\ref{CrI3_ml}(b). In absence of electric field, the CrI$_3$ monolayer is centrosymmetric, with $C_{3i}$ point-group symmetry (with $r_1$=$r_2$, $\theta_1$=$\theta_2$, and $d_1$=$d_2$ in Fig.\ref{CrI3_ml}). We have optimized its geometry and find the equilibrium lattice constant to be 6.99 \AA. The Cr-I bond distance ($r$) and Cr-I-Cr angle ($\theta$) are 2.75 \AA\ and 94.33$^o$, respectively. These values are consistent with previous DFT results\cite{C8TC01302K,Half-metallicity,PhysRevB_strain,chinese_PRB}.

From the spin-polarized DFT calculations we find that the FM state of CrI$_3$ monolayer is energetically more favorable than the AFM state by 18.36 meV/Cr, in good agreement with previous DFT results\cite{C8TC01302K,Half-metallicity}. The calculated atomic magnetic moment of Cr is 2.982 $\mu_\mathrm{B}$, close to the total magnetic moment of the FM monolayer per formula unit (3 $\mu_\mathrm{B}$), as observed previously\cite{C8TC01302K,Half-metallicity,PhysRevB_strain}.

\subsection{\bf{Modifications produced by the electric field}}

\begin{figure}[h]
\centering
\vspace{0cm}
{\includegraphics[height=0.81\textwidth]{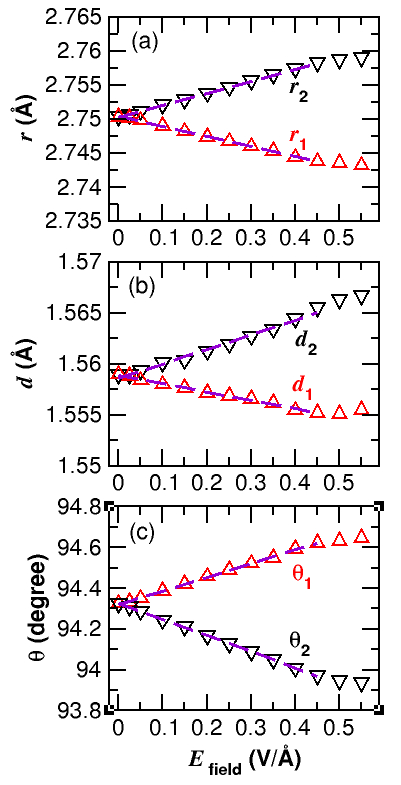}}
\caption{Structural parameters of CrI$_3$ as a function of the external electric field. (a) Bond distances $r_1$ (Cr-I in layer 1) and $r_2$ (Cr and I in layer 2). (b) Interlayer separations between I in layer 1 and Cr: $d_1$, and I in layer 2 and Cr: $d_2$. (c) Cr-I-Cr angles for the upper; $\theta_1$, and lower; $\theta_2$, halves of CrI$_3$ monolayer.}
\label{CrI3_plot}
\end{figure}

In Fig.~\ref{CrI3_plot} we show the effects of electric field on the CrI$_3$ geometry characterized by the Cr-I distance ($r$), difference in $z$ coordinates of Cr and I ($d$), and Cr-I-Cr angle ($\theta$) for I atoms from both layer 1 and layer 2 [see Fig.~\ref{CrI3_ml}(b)]. These parameters become inequivalent for the two I surface layers in the presence of the electric field, which breaks inversion symmetry. The splitting in the $\theta$ and $d$ values, i.e., $\Delta\theta$= $\theta_1$-$\theta_2$ and $\Delta d$=$d_2$-$d_1$, parameters, in particular, mostly control the Dzyaloshinskii-Moriya interaction between neighbouring Cr spins, responsible for the large ME effects predicted in
Refs.~\cite{chinese_PRB,chinese_skyrmion}.
The external electric field causes the I$^{-}$ ions to move along the direction opposite to the applied field, i.e., toward negative $z$-direction, while the Cr$^{3+}$ ions move along positive $z$-direction. Consequently, $r_1$, $d_1$ and $\theta_2$ decrease while $r_2$, $d_2$ and $\theta_1$ increase with $E_\mathrm{field}$. As can be seen in Fig.~\ref{CrI3_plot}, the linear-response behavior persists up to the $E_\mathrm{field}$ value of 0.45 V/\AA. Above that value, in the non-linear regime, the splittings $\Delta d$, $\Delta\theta$, and $\Delta r$=$r_2$-$r_1$ tend to saturate in Fig.~\ref{CrI3_plot}.

Our results in Fig.~\ref{CrI3_plot} show that the structural response of CrI$_3$ to the electric field up to 0.6 V/\AA\ is extremely small, for all structural parameters in the wide range of $E_\mathrm{field}$ values investigated. In particular, considering the $E_\mathrm{field}$ value of 0.2 V/\AA\ (which is the largest value examined in the previous theoretical work\cite{chinese_PRB,chinese_skyrmion} and is twice the maximal field value applied in the experimental study\cite{NatMaterJiang}), the resulting splitting in the bond distance $\Delta r$ [see Fig.~\ref{CrI3_plot}(a)] is 0.006 \AA\  (amounting to $\Delta r /r$=$0.2$ \%), the splitting in the interlayer distance  $\Delta d$ [Fig.~\ref{CrI3_plot}(b)] is 0.004 \AA\ ($\Delta d /d$=$0.3$ \%), and the splitting in the Cr-I-Cr angles $\Delta\theta$ [Fig.~\ref{CrI3_plot}(c)] is 0.29$^o$ ($\Delta\theta /\theta$=$0.3$ \%).  All of these structural changes amount to 0.3~\% or less. Such a very small impact of the applied electric field on the ionic structure is due, as we show in the Supplementary Material, to a very efficient electronic screening of the electric field within the CrI$_3$ monolayer.


As a result of the small structural changes induced by the electric field in Fig.~\ref{CrI3_plot}, we find that also the exchange energy, $\Delta E = E_\mathrm{AFM}-E_\mathrm{FM}$, of the CrI$_3$ monolayer barely changes when the electric field increases, e.g., at $E_\mathrm{field}$ = 0.2 V/\AA, $\Delta E = 18.34$ meV/Cr and at $E_\mathrm{field}$ = 0.45 V/\AA, $\Delta E = 18.24$ meV/Cr. The Cr magnetic moment of FM monolayer remains 2.982 $\mu_\mathrm{B}$ at  0.2 V/\AA\ and very slightly decreases to 2.978 $\mu_\mathrm{B}$ at 0.45 V/\AA. 


Our results are in contrast to the recent theoretical study reporting significant structural changes (up to 8~\% for $\Delta \theta / \theta$)\cite{chinese_PRB} in the linear regime for electric fields up to 0.2 V/\AA\ in the CrI$_3$ monolayer\cite{chinese_PRB,chinese_skyrmion}. As we show in the Supplementary Material, some non physical settings of the electric field result in such exaggerated (by a factor larger than 20) structural response to $E_\mathrm{field}$. The structural response we find is consistent, instead, with the negligible band-structure changes reported in another theoretical study induced by structural relaxation in a much larger electric field of 1 V/\AA\cite{C5CP04835D}.  
 
The Dzyaloshinskii-Moriya interaction, $\vec{D}$, between neighbouring Cr spins induced by the monolayer structural asymmetry or splitting $\Delta d$ (or equivalently by $\Delta \theta$) was found to be as large as $|\vec{D}| = 0.4$ meV (0.2 meV) for $E_\mathrm{field} = 0.1$~V/\AA\ (0.05~V/\AA) in 
Ref.~\cite{chinese_PRB}. 
Such large $|\vec{D}|$ values are comparable to the magnetic anisotropy energy (MAE) of the CrI$_3$ monolayer, i.e., 
0.8 meV/Cr\cite{PhysRevB_strain,chinese_PRB}, and therefore generated significant ME effects in the corresponding simulations\cite{chinese_PRB,chinese_skyrmion}, such as electrical reversal of magnetization\cite{chinese_PRB} and electric field induced magnetic skyrmions\cite{chinese_skyrmion}. When properly evaluated, however, the actual structural response to $E_\mathrm{field}$, in Fig.~\ref{CrI3_plot}, is one to two orders of magnitude smaller than in Refs.~\cite{chinese_PRB,chinese_skyrmion} (more than a factor 20 smaller than in Ref.~\cite{chinese_PRB}). 
The corresponding $|\vec{D}|$ magnitude, which scales linearly with $\Delta d$ ($\Delta \theta$) for such small distortions, is therefore less than 0.02 meV for $E_\mathrm{field} = 0.1$~V/\AA\ (less than $0.004$ meV for $E_\mathrm{field} = 0.02$~V/\AA) and is thus totally negligible compared to the CrI$_3$ MAE. Hence, in the linear regime (up to $E_\mathrm{field} = 0.45$~V/\AA), no significant structural distortion and no resulting relevant Dzyaloshinskii-Moriya ME effect can be reasonably expected to occur experimentally in the CrI$_3$ monolayer. 

We note that we have also examined the effect of the electric field on the atomic structure of the CrI$_3$ bilayer in the AFM configuration (with AA layer stacking). The PBE-D2 scheme was used to include dispersion interaction between the two layers of CrI$_3$\cite{Grimme}. The equilibrium separation between the CrI$_3$ layers was 3.44 \AA. Similar to the monolayer, no significant structural change was found in the bilayer CrI$_3$ due to the application of the electric field. The largest changes were of the same order of magnitude as for the monolayer. For example at $E_\mathrm{field}$ = 0.1 V/\AA, the interlayer distances $d_1$ and $d_2$ changed by 0.07\% and 0.10\% respectively for the upper monolayer, while for the lower monolayer $d_1$ and $d_2$ changed by 0.02\% and 0.08 \% respectively. Hence, as for the monolayer, the structural changes in the bilayer are far too small to induce any relevant Dzyaloshinskii-Moriya effect.

\label{}


\section{Conclusion}
In this study we have examined the effect of the external perpendicular electric field on the structural properties of the free-standing CrI$_3$ monolayer (as well as the bilayer).  We considered a wide range of electric field values, going beyond the linear-response regime, which we find to persist up to field values as large as 0.45 V/\AA. Above that value, the changes in the atomic structure associated with the field-induced breaking of inversion symmetry, tend to saturate. In contrast to previous theoretical findings, we show that the structural response to the electric field is far too small to induce any significant Dzyaloshinskii-Moriya magnetoelectric effect in the CrI$_3$ monolayer as well as in the bilayer. The experimentally observed large magnetoelectric response is thus bound to have a different microscopic origin.  



\newpage

\begin{flushleft}
{\textbf{\large{Associated Content}}}

{\textbf{Supporting Information}} Supporting Information accompanies this paper which is available free of charge.



{\textbf{Conflicts of interest}} 
The authors declare no conflict of interest for this work.

\newpage

\textbf{\large{Supplementary Material}}

\begin{figure}[b]
\centering
\vspace{0cm}
{\includegraphics[height=0.71\textwidth]{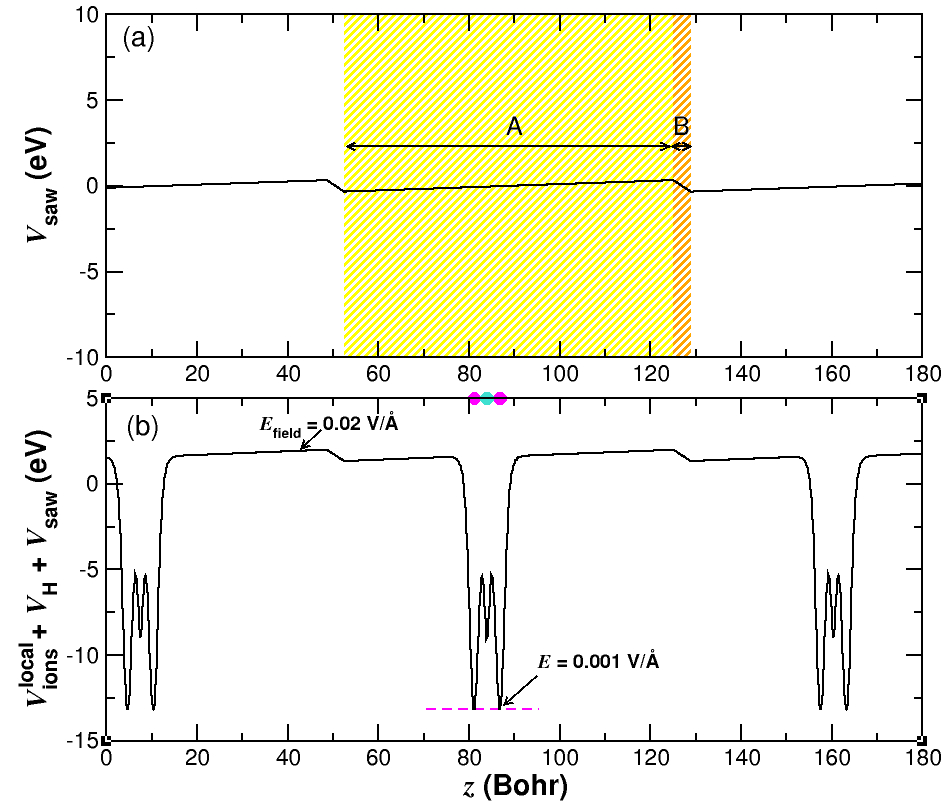}}
\renewcommand*{\thefigure}{S\arabic{figure}}\caption{Planar averaged plots of (a) sawtooth potential  $V_\mathrm{saw}$ and (b) $V_\mathrm{ions}^\mathrm{local}+V_\mathrm{saw}+V_\mathrm{H}$ for CrI$_3$ monolayer at $E_\mathrm{field}$ = 0.02 V/\AA\ for periodically repeated slabs. $V_\mathrm{ions}^\mathrm{local}$ is the linear superposition of the local part of the ions pseudo potentials, $V_\mathrm{saw}$ is the sawtooth potential, and $V_\mathrm{H}$ is the Hartree potential. In (a), region ``A'' shows the part of supercell where the external electric field is applied while the counterfield is present in region ``B''. The magenta and torquoise circles show the positions of the I and Cr atoms, respectively, along the $z$ direction. The electrons and ions of the CrI$_3$ monolayer strongly screen the external electric field, and as a consequence, potential of the iodine layers remain nearly unshifted in energy, as shown by the dashed magenta line in (b). The corresponding slope yields an electric field ($E$) value of 0.001 V/\AA\ within the CrI$_3$ monolayer.}
\label{E_p}
\end{figure} 

In Fig.~S\ref{E_p}(a) the sawtooth potential $V_\mathrm{saw}$ used in our calculations for the application of the external electric field, with $E_\mathrm{field}$ = 0.02 V/\AA, is plotted along $z$ direction, i.e., perpendicular to the plane of the CrI$_3$ monolayer (following the notation of Ref.~\cite{chinese_PRB}, the total potential acting on the electrons is $V_\mathrm{per}(\vec{r})+V_\mathrm{saw}(z)$). Fig.~S\ref{E_p}(b) shows the planar average of the corresponding self-consistent electrostatic potential: $V_\mathrm{ions}^\mathrm{local}+V_\mathrm{saw}+
V_\mathrm{H}$, where $V_\mathrm{ions}^\mathrm{local}$ is the local part of the linear superposition of atomic pseudopotentials, and $V_\mathrm{H}$ is the Hartree potential. In our calculations the external field, $E_\mathrm{field}$ is applied in region ``A'' (of length $l_\mathrm{A}$) while the resulting counterfield (to restore periodicity) is present in ``B''. The region ``A'' is centered around the monolayer and covers 95\% of the supercell. Region ``B'' lies in the vacuum region between the periodic images of the supercell and covers only 0.5\% of the supercell. From the calculations performed with the electric field $E_\mathrm{field}$ properly applied in the ``A'' region including the CrI$_3$ monolayer (illustrated in Fig.~S\ref{E_p}), we find extremely small changes in the ionic structure of the monolayer induced by the electric field, as reported in our paper.

In Fig.~S\ref{E_p}(b) one can observe that the potential minima corresponding to the positions of the two layers of the iodine atoms are nearly unshifted in energy with respect to each other compared to the shift $\Delta V$ that would be expected in the case of unscreened electric field, i.e., $\Delta V$ = $|{e}|E_\mathrm{field}\Delta z$, where $\Delta z$ is the iodine interlayer separation. The slope of the dashed line relating the two minima in Fig.~S\ref{E_p}(b) corresponds to a field $E = $ 0.001 V/\AA\ which is very small compared to the slope in vacuum region corresponding to the applied electric field $E_\mathrm{field}=0.02$ V/\AA. This indicates the presence of a large screening (electronic and ionic) of the electric field within the CrI$_3$ monolayer. The same behaviour is present in the calculated electrostatic potential also for larger applied electric field, for instance $E_\mathrm{field}=0.4$ V/\AA\ as shown in Fig.~S\ref{potential}. The slope of the dashed line corresponds to an electric field $E=0.02$ V/\AA\ within the CrI$_3$ monolayer which is considerably reduced with respect to the applied electric field. In fact, from the ratio of the electric field in the vacuum to the electric field in the monolayer, we estimate the dielectric screening (electronic+ionic) constant of CrI$_3$ would be $\epsilon_{0}$ $\approx$ 20. Similar calculations of the field ratio performed without ionic relaxation for the electronic screening alone gives $\epsilon_{\infty}$  $\approx$ 13. This large electronic screening of the electric field is the reason behind the very small displacements of the ions found in our calculations.


\begin{figure}[t]
\centering
\vspace{0cm}
{\includegraphics[height=0.71\textwidth]{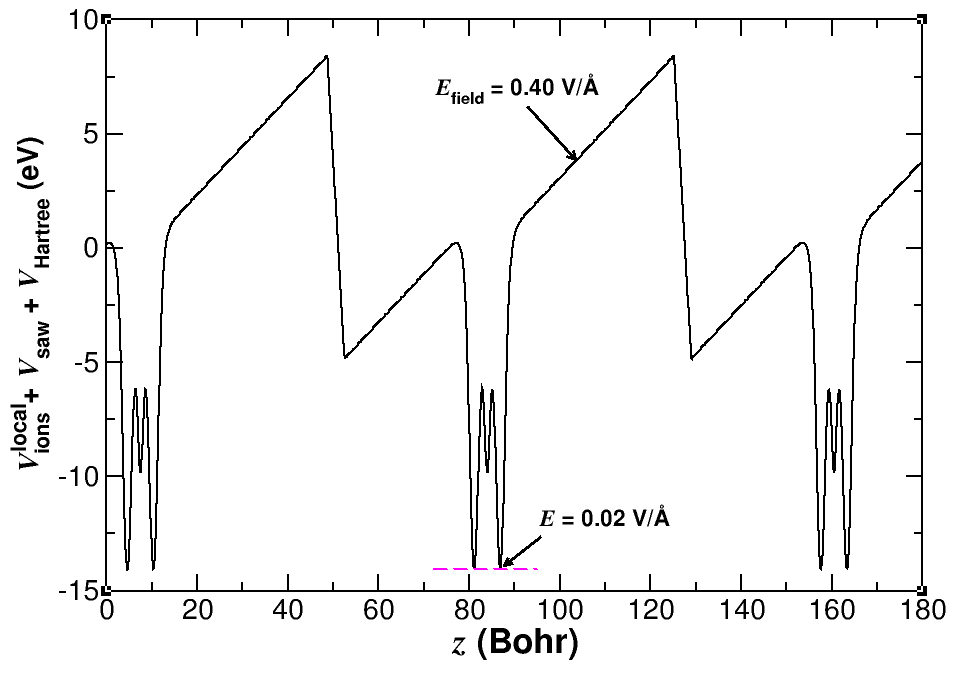}}
\renewcommand*{\thefigure}{S\arabic{figure}}\caption{Planar averaged plot of the potential ($V_\mathrm{ions}^\mathrm{local}+V_\mathrm{saw}+V_\mathrm{H}$) for CrI$_3$ monolayer at $E_\mathrm{field}$ = 0.4 V/\AA\ for periodically repeated slabs. $V_\mathrm{ions}^\mathrm{local}$ is the linear superposition of the local part of the ions pseudo potentials, $V_\mathrm{saw}$ is the sawtooth potential, and $V_\mathrm{H}$ is the Hartree potential. The electrons and ions of the CrI$_3$ monolayer strongly screen the external electric field and as a result of which the potential of the iodine layers remain nearly unshifted in energy, as shown by the dashed magenta line. The corresponding slope yields an electric field ($E$) value of 0.02 V/\AA\ within the CrI$_3$ monolayer.}
\label{potential}
\end{figure} 
 
If instead one inadvertently exchanges the location of the external electric field and of the counter field in the input of the sawtooth potential, i.e., puts the applied field $E_\mathrm{field}$ in region ``B'' (of the vacuum) resulting in a counter field present in region ``A'', then huge distortions in the structure of the CrI$_3$ monolayer are obtained. In the case when the external field is applied in region ``B'' of Fig.~S\ref{E_p}, which lies in the vacuum region, then one reproduces precisely all the results on the structural distortions as a function of the applied electric field reported in Phys. Rev. B, 97, 054416 (2018)\cite{chinese_PRB}. Using, e.g., $E_\mathrm{field}$ = 0.05 V/\AA, the structural changes are found to be: $\Delta d =  d_{1}-d_{2}=0.073$ \AA ($\frac{\Delta d}{d_{1}}$ of 4.8\%) and $\Delta \theta =  \theta_{1}-\theta_{2}=1.20^0$ ($\frac{\Delta \theta}{\theta_{1}}$ of 1.28\%). These values are more than one order of magnitude larger than values obtained with the proper physical setting of the field ($\Delta d = 0.001$ \AA\ and $\Delta \theta = 0.06^0$).

Such effect arises because the component due to the electric field of the forces ($F$) on the ions is calculated directly (analytically) from the value of the $E_\mathrm{field}$ according to Eq.~\ref{force}, below\cite{PhyReVBEfield}:
  
\begin{equation}
\vec{F}=\vec{F}^\mathrm{per}-eZ({\frac{4\pi m}{l_A}}-E_\mathrm{field})\cdot\hat{z} 
\label{force}
\end{equation}

where $F^\mathrm{per}$ is the Hellmann-Feynman force calculated with the potential satisfying the periodic boundary condition ($V_\mathrm{per}$)\cite{PhyReVBEfield}. The second part of Eq.~\ref{force} is evaluated analytically in terms of the applied field $E_\mathrm{field}$, the ion charge $Z$, and the dipole moment, $m$, of the slab\cite{PhyReVBEfield}. This part (which is substantial) is correct obviously only if the monolayer is inside the region of the applied field.



\end{flushleft}

\bibliography{CrI3_mono}
\bibliographystyle{spphys}

%
%

\end{document}